\documentclass[useAMS,usenatbib,12pt,psfig]{mn2e} 
\usepackage{color}
\usepackage{graphicx}
\usepackage{rotating}
\usepackage{txfonts}
\usepackage{pdflscape,lscape}

\def\hi{H\,{\sc i}}
\newcommand{\halpha}{H${\alpha}$}
\newcommand{\km}{km\,s$^{-1}$}
\newcommand{\msolar}{M$_{\odot}$}

\def\mh1{$M_{\rm H_{I}}$}

\newcommand{\prim}{$^{\prime}$}

\title{\hi\ in the Arp 202 system and its tidal dwarf candidate }
\author [Sengupta {\it{et al.}}]{Chandreyee Sengupta,$^{1}$\thanks{e-mail:sengupta.chandreyee@gmail.com(CS), t.c.scott2@herts.ac.uk(TS), dwaraka@rri.res.in(KSD), djs@ncra.tifr.res.in(DJS), bws@kasi.re.kr(BWS)} T. C. Scott$^{2}$, K. S. Dwarakanath$^{3}$, D. J. Saikia$^{4,5}$ and B. W. Sohn$^{1,6}$  \\ \\ 
$^{1}$ Korea Astronomy and Space Science Institute, 776, Daedeokdae ro, Yuseong gu, Daejeon, 305-348, Republic of Korea\\ 
$^{2}$ Centre for Astrophysics Research, University of Hertfordshire, College Lane, Hatfield, AL10 9AB, UK \\
$^{3}$ Raman Research Institute, Bangalore 560 080, India \\
$^{4}$ National Centre for Radio Astrophysics, Tata Institute of Fundamental Research, Pune 411 007, India \\
$^{5}$ Cotton College State University, Panbazar, Guwahati 781001, India \\
$^{6}$ Department of Astronomy \& Space Science, University of Science \& Technology, 217 Gajeong-ro, Daejeon, Korea \\}
\begin{document}

\date{Received  ; accepted  }
\date{}
\pagerange{\pageref{firstpage}--\pageref{lastpage}} \pubyear{}

\maketitle

\label{firstpage}

\begin{abstract}  We present results from our  Giant  Metrewave Radio Telescope  (GMRT) \hi\ observations of the interacting pair Arp 202 (\textcolor{black}{NGC\,2719 and NGC\,2719A}). \textcolor{black}{ Earlier  deep UV\textit{(GALEX)} observations of this system revealed a tidal tail like extension  with a diffuse object towards  its end, proposed as  a tidal dwarf galaxy (TDG) candidate.  We detect H{\sc i} emission from the Arp 202 system\textcolor{black}{, including  \hi\ counterparts for the tidal tail and the TDG candidate. Our  GMRT \hi\  morphological} and kinematic results clearly link the \hi\ tidal tail and the \hi\ TDG counterparts to the interaction between  NGC\,2719 and NGC\,2719A, thus strengthening the case for the  TDG. The Arp 202 TDG candidate  belongs to a small group of  TDG candidates with extremely blue colours. In order to gain a better understanding of this group we carried out a comparative study of their properties from the available data. We find that  \hi\ (and probably stellar) masses of this extremely blue group are similar to the lowest \hi\ mass TDGs in the literature.  However the number of such blue TDG candidates examined so far is too small to conclude whether or not their properties justify them to be considered as a subgroup of TDGs.}

\end{abstract}

\begin{keywords}
galaxies: spiral - galaxies: interactions - galaxies: kinematics and dynamics - 
galaxies: individual: Arp 202 - radio lines: galaxies \textcolor{black}{galaxies: tidal dwarfs}
\end{keywords}

\section{Introduction}
\label{intro}

Tidal interactions between galaxies, where at least  one of them is gas rich,  can result in tidal stripping of large amounts of  \hi\ from the potential of the parent galaxy(s). Most of this stripped \hi\  will eventually fall back into the potential of one or other of the interacting galaxies or be  incorporated  into the intra--group medium (IGM). But  if the \hi\  densities are sufficient and environmental conditions are favourable, self-gravitating bodies with masses typical of dwarf galaxies, called Tidal Dwarf Galaxies (TDG), may form within the tidally stripped gas  (Duc \& Mirabel 1999; Duc et al. 2000). \textcolor{black}{Apart from establishing that these are indeed self-gravitating objects,} a key observational problem is \textcolor{black}{to distinguish} TDGs from  older standard dwarfs because some TDGs contain old stars \textcolor{black}{stripped} from their parent galaxies. \textcolor{black}{ Other TDGs} consist almost entirely  of young stars formed in situ  from \hi\ stripped from the parents during the interaction (Duc \& Mirabel 1999; Duc et al. 2000; Braine et al. 2001). A combination of evidence linking the TDG candidate to  interacting parents and stellar population  studies is normally used to identify TDG candidates, although indisputable  criteria remain to be accepted.    

TDGs are usually observed during an active star forming stage thus allowing tests of star formation criteria and stellar to gas relationships in low gas density environments. As TDGs form from stripped gas they are expected to contain little or no dark matter \textcolor{black}{(Elmegreen et al. 1993)}.  This  makes them valuable in studying the role and importance of dark matter in galaxy formation.  TDGs are also expected to be metal rich in comparison to  normal dwarfs, \textcolor{black}{because} the gas from which TDGs form originates from more evolved, usually large, late--type galaxies. The lack of  dark matter in \textcolor{black}{TDG's implies} they are more vulnerable to disruption by further tidal interactions than normal dwarfs. \textcolor{black}{Thus} TDGs  provide a unique environment to investigate the processes governing  formation and evolution of galaxies.

\onecolumn

\begin{landscape}
\begin{figure*}
\centering

\includegraphics[scale=0.78, angle=0]{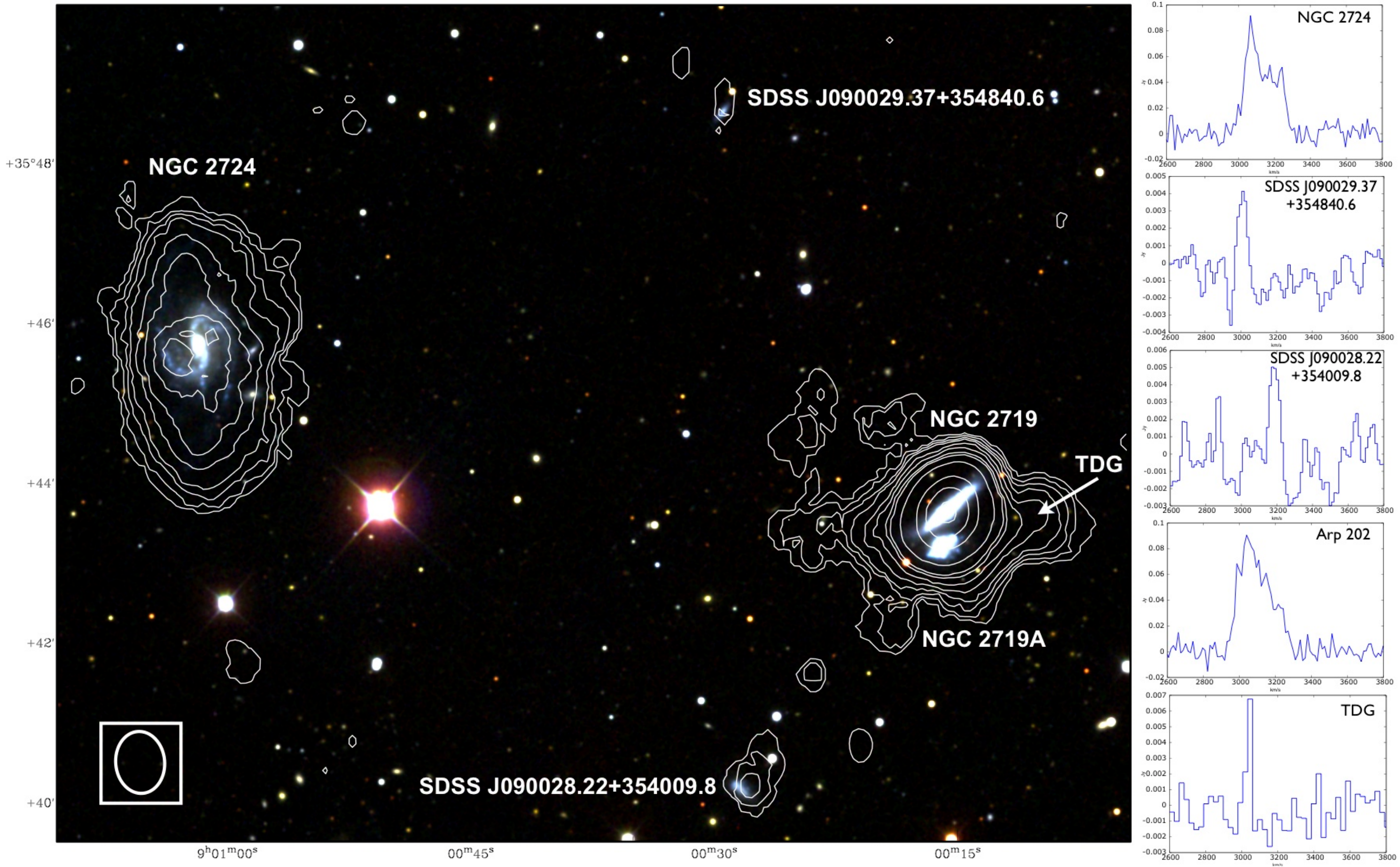}  
\caption{\textcolor{black}{ Arp\,202 field: \textcolor{black}{Left: Low-resolution  integrated \hi\ emission contours overplotted on a SDSS \textit{g, r, i}--band false colour image.  The contours are  \hi\ column
density levels of 10$^{19}$ atoms /cm$^{2}$ x (4.4, 9.3, 15.5, 21.7, 34.1, 46.5, 61.9, 71.2, 105.3, 185.8, 247.7, 309.7). \textcolor{black}{The first contour is at 3 $\sigma$.}  The beam size (47.5$^{\prime\prime}$  x 37.4$^{\prime\prime}$) is indicated by the ellipse at the bottom left of the image. Names of galaxies detected in H{\sc i}} are shown adjacent to their optical positions. \textbf{Right:} GMRT \hi\ spectra, with a velocity resolution of 27 \km, for each of the detections. The Arp 202 spectrum \textcolor{black}{is from the integrated} emission from NGC\,2719, NGC\,2719A, the tidal tail and the TDG candidate. }  } 
\label{fig1}

\end{figure*}
\end{landscape}

\twocolumn

To address the issue of in--situ star and TDG formation in tidal gas debris, a multi--wavelength study of a sample of Arp interacting galaxies (the ‘Spirals, Bridges, and Tails’ (SB\&T) sample) is being carried out \citep{Smith, smith2010}. \textit{GALEX}\footnote{Galaxy Evolution Explorer}, Spitzer and H$\alpha$ observations of these systems have revealed evidence of in-situ star formation and TDG candidates. One such system is the Arp 202  close pair of galaxies, NGC\,2719 and  NGC\,2719A, which have  radial velocities 3077 km s$^{-1}$ and 3117 km s$^{-1}$ respectively and  signs of a tidal interaction at optical \textcolor{black}{and UV} wavelengths (Figures \ref {fig1} and \ref{fig2}). Deep UV \textit{(GALEX)} observations of this system revealed a tidal tail like extension which appears to originate from NGC\,2719A with a diffuse object towards  its end, \textcolor{black}{proposed as  a TDG by \cite{smith2010}}.  

In this paper, we present results from our  Giant  Metrewave Radio Telescope  (GMRT) \hi\ observations of the Arp 202 system including at the position of the proposed TDG. The paper also utilises \textcolor{black}{publicly} available \textcolor{black}{SDSS\footnote{Sloan Digital Sky Survey}, \textit{Spitzer}  and \textit{GALEX}\footnote{Galaxy Evolution Explorer}}  data and images.  Section \ref{obs} sets out details of our observations, and the results are given in section \ref{results}. We discuss the results in section \ref{dis}. A summary and concluding remarks are set out in section \ref{summary}. The average of the radial velocities of NGC\,2719 and NGC\,2719A is 3097 km s$^{-1}$. Using this average velocity and assuming H$_{0}$ to be 75 km s$^{-1}$ Mpc$^{-1}$, we adopt a distance of 41.3 Mpc for NGC\,2719 and  NGC\,2719A and the TDG. At this distance the spatial scale is 12 kpc / arcmin. J2000 coordinates are used throughout the paper, including in the figures.

\section{Observations}
\label{obs}
\hi\  observations of Arp 202 were carried out with the GMRT on April 30th, 2008. Further details of the observations  are given in Table \ref{table1}. The baseband bandwidth used was 8 MHz for the \hi\ 21-cm  line observations giving a velocity resolution of $\sim$13 km~s$^{-1}$. 

The  Astronomical Image Processing System ({\tt AIPS}) software package was used for data reduction. Bad data  from malfunctioning antennas, antennas with low gain and/or radio frequency interference (RFI) were flagged.   The flux densities are on the scale of \cite{baar}, with flux density uncertainties of  $\sim$5\%.  \textcolor{black}{ Following calibration, continuum subtraction was carried out  in the \textit{uv} domain } using the {\tt AIPS} tasks  \textsc{uvsub} and \textsc{uvlin}. To analyse the  \hi\ structures,  image cubes of different resolutions were produced by applying different `tapers' to the data with varied \textit{uv} limits. \textcolor{black}{The task \textsc{imagr} was then used to obtain  the final cleaned  \hi\  image cubes. The integrated \hi\  and  \hi\   velocity field maps were extracted from the cubes using the AIPS task \textsc{momnt}. } Details \textcolor{black}{ for} the final high and low-resolution cubes are given in Table \ref{table1}.

\begin{table}
\centering
\begin{minipage}{110mm}
\caption{GMRT observation details}
\label{table1}
\begin{tabular}{ll}
\hline


Frequency & 1420.4057 MHz \\
Observation Date &30th April, 2008 \\
Primary calibrator&3C147\\ 
Phase Calibrator & 0741+312 (2.00 Jy)  \\
(flux density)  &   \\
Integration time  & 8.0 hrs  \\
primary beam&24\arcmin ~at 20 cm \\
Low resolution beam & 47.5$^{\prime\prime}$ $\times$ 37.4$^{\prime\prime}$  (PA = 5.9$^{\circ}$) \\
High resolution beam & 23.4$^{\prime\prime}$ $\times$ 16.3$^{\prime\prime}$ (PA = 17.8$^{\circ}$) \\

rms for low-resolution map  & 1.24 mJy beam$^{-1}$  \\
rms for high-resolution map & 0.81 mJy beam$^{-1}$  \\

RA (pointing centre)& 09${\rm h}$ 00${\rm m}$ 17.17${\rm s}$ \\
DEC (pointing centre)& 35$^\circ$ 43$^\prime$ 33$^{\prime\prime}$.12\\

\hline
\end{tabular}
\end{minipage}
\end{table}

\section{Observational Results} 
\label{results}
 
\begin{figure*}
\centering
\includegraphics[scale=0.7, angle=0]{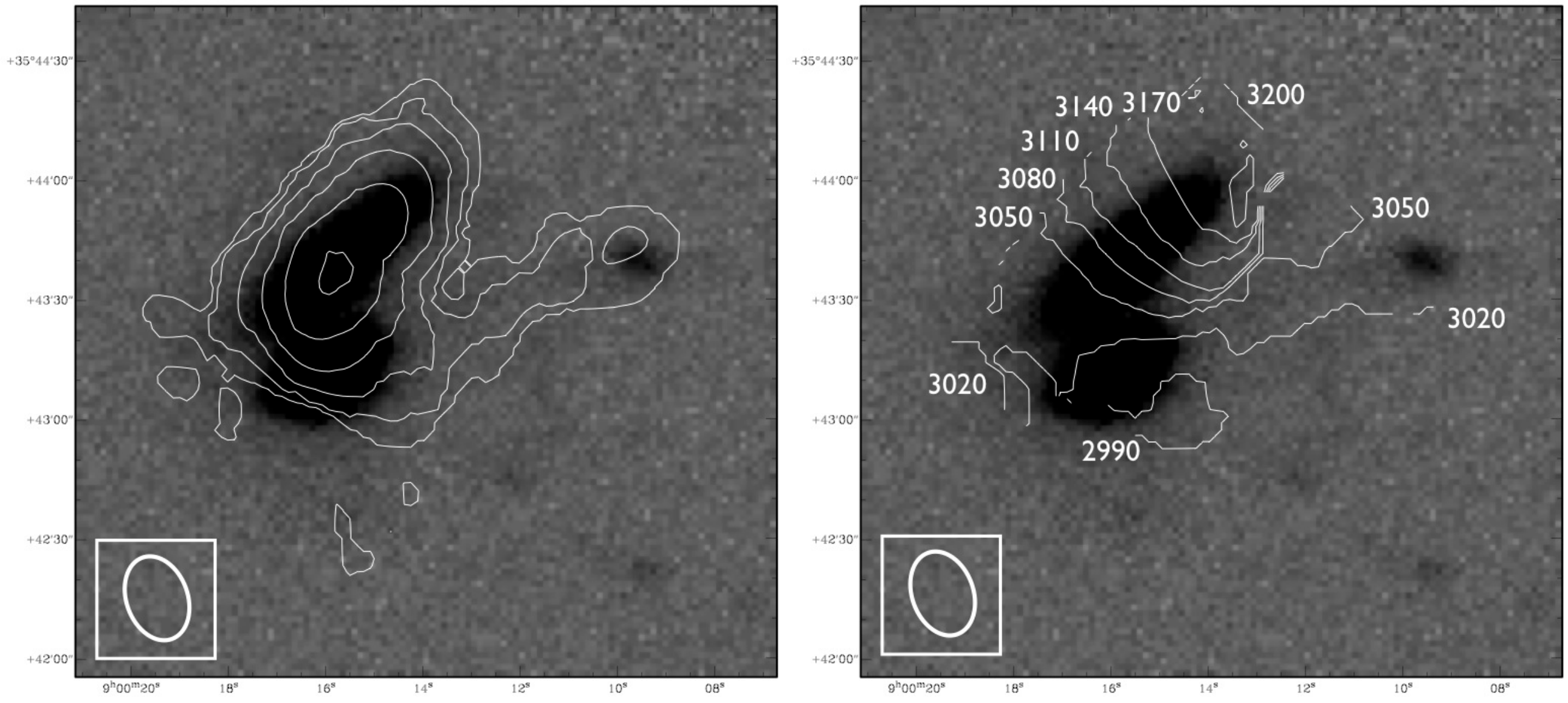}  
\caption{\textcolor{black}{Arp\,202: \textbf{ Left panel } High resolution  \hi\ integrated map contours (white) overlaid on a NUV \textit{(GALEX)} image.  The \hi\ column density contour levels are  10$^{20}$ atoms /cm$^{2}$ x (4.4, 7.5, 14.5, 26.0, 43.4, 66.6) .   \textbf {Right panel } \hi\ velocity  with contours at  velocity separation of 30\km.
The \hi\ beam (23.4$^{\prime\prime}$  x 16.3$^{\prime\prime}$) is shown  as white ellipse  at the bottom left of each panel.  } } 
   \label{fig2}
  \end{figure*}

\begin{figure*}
\centering
\includegraphics[scale=0.80, angle=0]{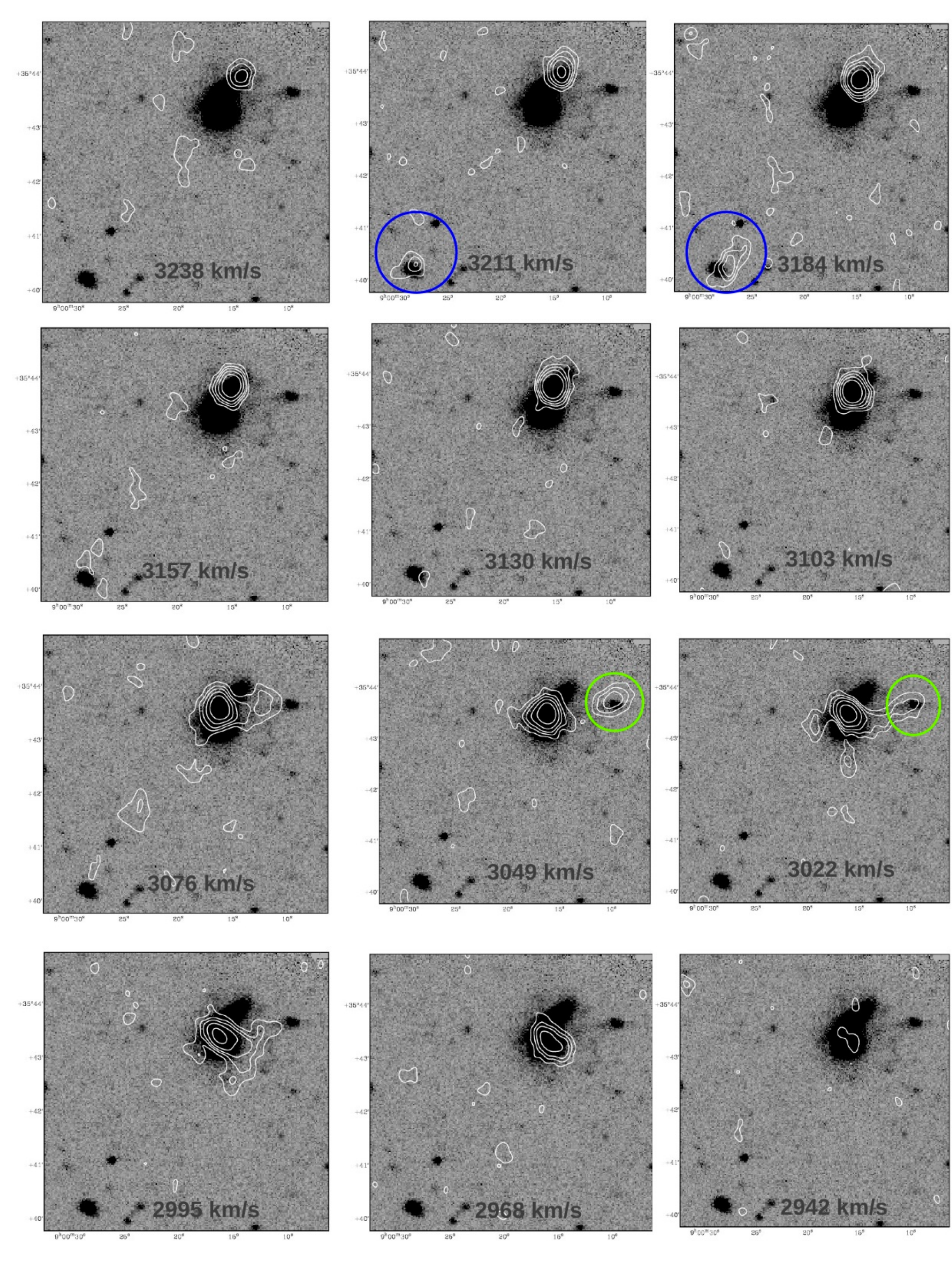}  
\caption{Channel images from the high resolution cube (beam is  23.4$^{\prime\prime}$ $\times$ 16.3$^{\prime\prime}$) showing the Arp 202 pair, the candidate TDG (indicated by green circle) and the fragmented H{\sc i} bridge between Arp 202 and SDSS J090028.22+354009.8 (indicated by blue circle). The contour levels are 0.7 mJy $\times$ (3,5,7,10,15). \textcolor{black}{The H{\sc i} contours are overlayed on the SDSS r-- band image}. } 
   \label{fig4}
  \end{figure*}

\textcolor{black}{The \textcolor{black}{contours in Figure \ref{fig1} are} from the  low-resolution (47.5$^{\prime\prime}$ x 37.4$^{\prime\prime}$) integrated \hi\ GMRT map for the Arp 202 field  overlaid on a SDSS \textit{g, r, i}--band false colour image.}  \hi\ properties of the objects  for the Arp\,202  field are  set out   in Table  \ref{table2}. The two  principal \hi\ detections are, the large SABc spiral,  NGC\,2724,  in the north-east  and  the Arp\,202 system in the west detected in the velocity range 2942 \km\ to 3238 \km. Within the Arp\,202 system the main  \textcolor{black}{features are the  \hi\ counterparts of  the NUV \textit{(GALEX)}  } tidal tail and TDG candidate. Figure \ref{fig2}  shows the high-resolution  (23.5$^{\prime\prime}$ $\times$ 16.3$^{\prime\prime}$) \hi\ integrated map contours overlaid on a NUV (\textit{GALEX}) image of the two principal galaxies in Arp\,202 system and the TDG candidate.  Figure \ref{fig2}, which shows  the tidal tail -- TDG connection at \hi\ column densities \textcolor{black}{$\gtrsim$} 4.3$\times$10$^{20}$ cm$^{-2}$. The peak \textcolor{black}{\hi\ column density found at the position of the TDG candidate is $\sim$ 7.5$\times$10$^{20}$ cm$^{-2}$}.  In the channel images (Figure \ref{fig4}) the \hi\ tail is visible from velocities 2995  \km\ to 3076  \km\ and the \hi\ peak emission in the tail coincides with the TDG in the velocity range of 3022  \km\ to 3049  \km.  The projected length  of the \hi\ tidal tail \textcolor{black}{which} appears to originate from NGC\,2719A,   is $\sim$  1.5$^{\prime}$ (20 kpc).
\begin{table*} 
\centering
\caption{GMRT \hi\ detections}
\label{table2}
\begin{tabular}[h]{@{}lllrrr@{}}
\hline
Object& RA& Dec&Velocity& M(\hi) \\
      &   &    & \hi\ &   x 10$^9$\\
&[h:m:s]&[d:m:s]  &[\km]&[\msolar]  \\
\hline
Arp 202 pair:& & &&7.1 $\pm$ 0.3\\
-- NGC\,2719&09:00:15.4&+35:43:40\\
-- NGC\,2719A&09:00:15.9&+35:43:12\\
-- TDG candidate&09:00:09.3&+35:43:38&3047 $\pm$ 13.6 & 0.1 $\pm$ 0.05\\
NGC\,2724&09:01:01.8&+35:45:43 &3220 $\pm$ 13.6& 5.6 $\pm$ 0.4\\
SDSS J090029.37+354840.6&09:00:29.4&+35:48:41&3081  $\pm$ 13.6 & $\le$ 0.1\\
SDSS J090028.22+354009.8&09:00:28.2&+35:40:10&3301  $\pm$ 13.6 & $\le$ 0.1\\
\hline

\end{tabular}

\textcolor{black}{{\small Uncertainties on H{\sc i} masses quoted here are for an adopted distance and inclination angle,\\
 estimated using uncertainty on the line flux. Actual uncertainties on the masses can be higher.}}
\end{table*}

Apart from the galaxies referred to in the previous paragraph, two further galaxies were detected in \hi\ within the FWHM of the GMRT primary beam:  SDSS J090028.22+354009.8 projected  $\sim$ 4.1$^{\arcmin}$ to the south-east of the Arp 202  and SDSS J090029.37+354840.6 projected  $\sim$ 5.9$^{\arcmin}$ to the north-east of Arp\,202. SDSS J090029.37+354840.6 was marginally detected (Figure 1), however the spectrum and channel maps show  the H{\sc i} signal to be correlated in consecutive channels at the optical position of the galaxy. We therefore conclude the \hi\ is associated with optical galaxy.

The GMRT  spectrum for each of the galaxies \textcolor{black}{detected in \hi\ and the Arp\,202 system } is  plotted \textcolor{black} {on  the right side of Figure \ref{fig1}. }The GMRT integrated flux density from the Arp 202 system is 17.55 Jy \km, which compares well with the uncorrected single dish  flux density of 16.70 Jy km$s^{-1}$ for NGC\,2719 \citep{hucht}.  The H{\sc i} mass derived from the GMRT  for the Arp\,202 pair, tidal tail  and  the candidate TDG is 7.1$\pm$ 0.3 $\times$10$^{9}$ M$_{\odot}$. We extracted an \hi\ spectrum from within approximately a 23 by 16 arcsec region (similar to the beam) centred on the TDG (Figure 2)  and derived an \hi\ mass \textcolor{black}{for the TDG} of  $\sim$ 1 $\pm$ 0.5 $\times$10$^{8}$ \msolar. The integrated flux density from the GMRT spectrum of NGC\,2724 is 13.9 Jy km$s^{-1}$ which compares well with the single dish value of 13.8 Jy km$s^{-1}$ \citep{2005ApJS..160..149S}. 
SDSS J090028.22+354009.8, which is $\sim$ 49 kpc to the south--east of the Arp pair, shows \textcolor{black}{signatures of interaction in H{\sc i}  maps} and is connected to the pair by a very faint and fragmented H{\sc i} bridge. The bridge can be traced in the channel images (Figure \ref{fig4}), although the signal to noise ratio is poor. 

\textcolor{black}{Figure \ref{fig2} -- right panel shows the velocity field of the Arp\,202 system from the high-resolution cube. The contours between 3050 \km\ to 3200 \km\ indicate the \hi\ in the central and NW parts of the NGC\,2719 disk are in a regularly rotating  edge on disk, although with indications of a warp. This rotation pattern is broken at south-eastern end of the  NGC\,2719 \hi\  disk  at velocities below 3050 \km. The abrupt change of rotation pattern in the contours together with the tidal tail morphology at velocities below 3050 \km\ indicates the \hi\ there has been strongly perturbed by  the interaction between NGC\,2719 and NGC\,2719A. The narrow range of  \hi\ velocities, $\sim$ 60 \km, in the tidal tail running westward from NGC\,2719A to and including  the TDG candidate  provide clear evidence linking the both the tidal tail and TDG candidate  to the interaction between NGC 2719 and NGC 2719A.}


\section{Discussion}
\label{dis}



\subsection {Arp 202 -- major interactions} The   pre-merger interacting pair\textcolor{black}{, NGC\,2719 and NGC\,2719A,} have  an optical  radial velocity separation of 40 \km\ and D$_{25}$  of 1.3 arcsec (17.5 kpc) and 0.9 arcsec (12 kpc) respectively. We estimate  the stellar mass  for NGC\,2719 at $\sim$  3.4 x 10$^9$ \msolar\  and NGC\,2719A  at   5.0 x 10$^9$ \msolar\  (a stellar mass ratio of $\sim$ 1:1.4), based on the parameters from \cite{Bell2003}  and \cite{Blanton} for SDSS \textit{g} --band magnitudes and \textit{g--r} colours. The sizes, SDSS colours and stellar masses indicate these are relatively small late--type galaxies (Im pec per NED\footnote{NASA/IPAC Extragalactic Database.}).  In optical images  NGC\,2719 presents as an essentially intact  high inclination disk.   While both galaxies are designated as late--types, the high-resolution \hi\ channel maps in Figure \ref{fig4}  show that almost all of the  \hi\ emission is projected against the NGC\,2719 disk.  Despite NGC\,2719 having the optical appearance of an  edge--on spiral, the GMRT \hi\ spectrum lacks double horned structure expected from a rotating \hi\ disk. This and the  tidal perturbation parameter\textcolor{black}{\footnote{$p_{gg}= \frac{(M_{comp} / M_{gal}) }{(d/r_{gal})^3} $  \textcolor{black}{\citep{param}}, where \textcolor{black}{$M_{gal}$ (5 $\times$ 10$^9$ \msolar) and $M_{comp}$ (3.4 $\times$ 10$^9$ \msolar)  are the masses of the galaxy  and companion respectively, $d$ is the separation (0.461\prim) and $r$ is the galaxy disk optical radius (0.45\prim). Values} of $p_{gg} > 0.1$ are likely to lead to tidally induced star formation.}, $p_{gg}$ = 0.6}, suggests  the NGC\,2719 \hi\ disk may have been severely perturbed by the interaction between the pair.   NGC\,2719A has a more irregular optical morphology, and the UV tail (Figure \ref{fig2}) appears  to originate from this galaxy.  It seems likely the \hi\ and stellar material in \textcolor{black}{the} tidal tail has its origin in a close interaction between NGC\,2719 and NGC\,2719A, where NGC\,2719A has lost \textcolor{black}{most of  its \hi\ } to NGC\,2719 and the IGM. 


The Arp\,202 pair are  H{\sc i} rich, with a combined  \hi\ mass of 7.1 $\times$ 10$^{9}$ M$_{\odot}$. \textcolor{black}{Galaxies of same size and morphological type as NGC\,2719 and 2719A are  expected to have   H{\sc i} masses  of $\sim$ 2.8 $\times$ 10$^{9}$ M$_{\odot}$ and $\sim$ 1.7$\times$10$^{9}$ M$_{\odot}$ respectively \citep{1984AJ.....89..758H}, making a total of 4.5$\times$10$^{9}$ M$_{\odot}$ for the pair.}. The presence of excess H{\sc i} mass \textcolor{black}{may be due to  the system acquiring} \hi\ though interactions and/or accretion of small neighbours. A 60$^{\arcmin}$ radius (720 kpc at that distance) search around the Arp 202 system, within a velocity range of 2000 km s$^{-1}$ to 4000 km s$^{-1}$ , yields only one similar sized neighbour (NGC 2724) and two small dwarfs (SDSS J090028.22+354009.8 and SDSS J090029.37+354840.6). The projected  distance between Arp\,202 and NGC 2724 is 9.6$^{\arcmin}$ (115 kpc). Assuming \textcolor{black}{galaxies are moving at$\sim$ 150  km s$^{-1}$, a typical number for velocity dispersion in loose groups,} we estimate that if they have crossed paths, then the last interaction could have \textcolor{black}{occurred}  about 3.6 $\times$ 10$^{8}$ years ago, less than the dynamical time scales of NGC\,2719 and NGC 2724, which are estimated to be 4.2 $\times$ 10$^{8}$ years and 5.2 $\times$ 10$^{8}$ years respectively. Had there been an interaction involving a massive transfer of \hi\  between these two galaxies, extended tidal features would have remained in  the \hi\ disks for at least one dynamical timescale. While the \hi\ in Arp 202 is heavily perturbed, the NGC\,2724  H{\sc i} disk is relatively undisturbed and shows no significant \hi\ deficiency. The expected H{\sc i} mass of NGC 2724 is 2.9 $\times$ 10$^{9}$ M$_{\odot}$  \citep{1984AJ.....89..758H} compared to the observed mass of 5.6 $\pm$ 0.4 $\times$ 10$^{9}$ M$_{\odot}$, making it an H{\sc i} rich spiral. Thus even if it has interacted with the Arp 202 system, it seems unlikely that any massive gas exchange has taken place between them. We conclude that  the perturbation signatures in the Arp\,202 are primarily the result of the interaction between NGC\,2719 and 2719A.

\subsection {SDSS J090028.22+354009.8 and SDSS J090029.37+354840.6 -- minor interactions} Apart from NGC\,2724, there are two dwarfs close to the Arp 202 system. SDSS J090028.22+354009.8, to the south-east is $\sim$ 4.1$^{\arcmin}$ (49 kpc) away and SDSS J090029.37+354840.6 to the north east is $\sim$  5.8$^{\arcmin}$ (69 kpc) away. Their stellar masses were estimated from the SDSS photometric measurements quoted in NED and using the mass to light ratio to colour relation prescribed in \cite{Bell2003}. The value for solar absolute magnitude in \textit{g} -- band was taken from \cite{Blanton}. The stellar masses for  SDSS J090028.22+354009.8 and SDSS J090029.37+354840.6 are estimated as 3.9 $\times$ 10$^{9}$ M$_{\odot}$ and   1.1 $\times$ 10$^{8}$ M$_{\odot}$ respectively.  \textcolor{black}{ Their H{\sc i} masses \textcolor{black}{from the GMRT data are} estimated to be $\le$ 1$\times$ 10$^{8}$ M$_{\odot}$. The upper limit was applied as they both are not unambiguously three sigma detections}. Both the galaxies are \textcolor{black}{beyond the fields imaged by } \textit{Spitzer} and only SDSS J090028.22+354009.8 is within the \textit{GALEX} Medium Imaging Survey (MIS) field. The galaxy is bright in the \textit{GALEX }map indicating  star formation activity within the last 10$^8$ yr. This is not unexpected as  SDSS J090028.22+354009.8 shows signs of recent interaction  with the Arp\, 202 system. We have detected fragments of an \hi\ bridge connecting Arp\,202 and SDSS J090028.22+354009.8. We see no similar H{\sc i} connection between Arp\,202 and SDSS J090029.37+354840.6. But we notice an offset between the optical and the H{\sc i} radial velocities for both these galaxies. While SDSS J090028.22+354009.8 and SDSS J090029.37+354840.6 have optical velocities of 3301  km s$^{-1}$ and 3081  km s$^{-1}$ in NED, the H{\sc i} is detected at velocities $\sim$  3181 km s$^{-1}$ and 3011  km s$^{-1}$ respectively. The H{\sc i} velocities quoted here correspond to the peak intensities. The H{\sc i} line widths at 20\% of the peak flux value of SDSS J090028.22+354009.8 and SDSS J090029.37+354840.6 are 109  km s$^{-1}$ and 82  km s$^{-1}$ respectively.  This shift of the H{\sc i} velocity could be due to  interactions with the Arp 202 system, during which the gas disks of the smaller galaxies have been disturbed. \textcolor{black}{Also \hi\ loss from these galaxies to}  the Arp\,202 system cannot be ruled out.

Since it is difficult to establish if such nearby dwarfs are indeed TDGs, we compared the available information on these galaxies to those of TDGs and probed the  chances of them being \textcolor{black}{Arp 202 system} detached TDGs. Judging by the optical and H{\sc i} masses and the SDSS g--r colours, SDSS J090028.22+354009.8 seems to be an old dwarf galaxy associated with the Arp 202 system. Its g--r colour is an extreme 1.65, much higher than usual early type dwarfs making it an  unreliable number to derive concrete conclusions.  SDSS J090029.37+354840.6 on the other hand is a bluer dwarf, with g-r value as 0.19, which is considered low in  dwarf galaxy samples \citep{barazza} and similar to blue compact dwarf galaxies (BCDG) \citep{noah}. In terms of stellar populations, tidal dwarf galaxies are known to be of two types, one with extremely new stars, with optical colours similar to BCDGs and the other dominated by an older population, mostly derived from the fragments of the parent galaxies \citep{ducmirabel1999}. However the dynamical mass to gas mass ratio of SDSS J090029.37+354840.6 is $\sim$8. \textcolor{black}{ Because of the predicted  absence of dark matter, this ratio is expected to be closer to 1 in TDGs}. Additionally,  the galaxy is about 70 kpc away from the main system and previous studies about TDG detachment shows that 95\% of TDGs are found within 20 kpc of their parent system \citep{kaviraj}. Thus from the information available \textcolor{black}{we conclude  these dwarfs are not detached TDGs, but rather} old satellites of the system. Metallicity and CO data  for  these dwarfs could \textcolor{black}{further} constrain the origin of these  dwarfs.

\subsection {The candidate TDG of the Arp 202 system}  \cite{Smith2010b}  first reported  the presence of a  clumpy star forming tidal tail terminating in \textcolor{black}{the TDG candidate from  NUV (\textit{GALEX}) observations  and  confirmed the candidate was part of the Arp 202 system with optical spectroscopy. Our GMRT \hi\  morphology and kinematic results, set out in section  \ref{results},  clearly show that the  \hi\ tidal tail/bridge  links the  \hi\ TDG counterpart  to the interacting Arp 202 pair. }  

The tail and TDG candidate are  faintly visible in individual SDSS images but \textcolor{black}{are much more apparent in the UV \textit{(GALEX) }images.  Neither were}  detected in either the Spitzer 8 $\mu$m or SARA \halpha\ images. This lead \cite{Smith2010b} to  suggest that the TDG may be in a post--starburst stage , i.e. dominated by a stellar population with age between  10$^7$ yr to 10$^8$ yr.  \cite{ngc4656} presented a \textit{FUV--g} vs \textit{g--r}  plot (their figure 11)  for  the  UV  detected TDG  candidates from the \cite{Smith2010b} \textcolor{black}{ and two other candidates, Holmberg IX and NGC 4656UV.} \textcolor{black}{The Arp 202 TDG candidate appears in their plot as one of the  candidates with extreme blue colour, together with  NGC\,4656UV, Arp\,305 and Holmberg IX, one of the strongest TDG candidates \citep{sabbi}. The mean  \textit{FUV--g}  colour of these four TDG candidates \textcolor{black}{(blue candidates)} is close to 0,  $\sim$ 1 mag \textcolor{black}{lower} than the rest of the  \cite{Smith2010b}  sample.  \textcolor{black}{But as  \cite{ngc4656} point out  a larger sample is needed to determine whether the blue candidates are part of a separate category of TDGs. While acknowledging this limitation, in Table \ref{table3}, we compare selected properties of the Arp\,202 TDG candidate  with those of the other three blue candidates and statistics from larger samples of TDG candidates \citep{kaviraj, ducmirabel1999}. The aim of compiling the data in the table was to determine if the blue candidates share any  additional distinguishing features. In the following paragraphs we briefly comment sequentially on each row in Table \ref{table3}.  }} 

\textcolor{black}{In  Table  \ref{table3}  the  mass and size estimates, in particular, are highly uncertain, so we are only able to draw order of magnitude conclusions from their comparative data. The uncertainties are particularly acute for the estimates of stellar mass (M*). M* in units of 10$^8$ \msolar\ for Arp\,202 (0.2) and  NGC\,4656UV (0.4) were estimated using  \textit{g} and \textit{r} band colours following the method described in note a to Table \ref{table3}. M* for Holmberg XI (0.02) and Arp\,305 (0.04) are taken from the literature (see  Table  \ref{table3} for references) and they were determined from SED fitting in the respective papers.  For Arp\,305 M* determined with the  \textit{g} and \textit{r} band colour method is 1.1 $\times$ 10$^8$ \msolar\   suggesting  this method may overestimate M* by at least an order of magnitude.  The \textit{g} and \textit{r} colour M* method assumes the galaxy has a standard mix old and young stellar populations but we have evidence from the non--detections of the Arp\,202 TDG candidate in Spitzter 3.6, 4.5 and 8 $\mu$m images (and only marginal equivalent detections for Arp\,305) that these two candidates are relatively deficient in old stellar populations compared to standard galaxies. }

\textcolor{black}{From the GMRT spectrum  for the  Arp\,202 TDG candidate (Figure \ref{fig1} -- Bottom right panel)  we estimate its H{\sc i} mass to be $\sim$ 1.0 $\times$ 10$^{8}$ M$_{\odot}$. \textcolor{black}{The mean M$_{HI} $ of the blue candidates is \textcolor{black}{ 2.2} $\pm$1.6 $\times$ 10$^{8}$ \msolar\ which is  an order of magnitude lower than  the mean value from [n=20] }\cite{ducmirabel1999}  sample (16 x 10$^{8}$ \msolar).  M$_{HI}$  for all the blue candidates is close to the minimum M$_{HI}$ of 2.0 $\times$ 10$^{8}$ M$_{\odot}$ in \cite{ducmirabel1999} sample. So even allowing for the uncertainties in the estimates of  M$_{HI}$, the blue candidates have a significantly lower M$_{HI}$  than a typical TDG candidate. Moreover if our interpretation of stellar mass in the previous paragraph is correct then Arp\,202 probably shares a M$_{HI}$ / M* ratio, of several tens to a hundred or so, with the other blue candidates.}

\textcolor{black}{For the  Arp\,202 TDG candidate,  M$_{dyn}$ is estimated at  3.9 $\times$ 10$^8$  \msolar\ using the method described in Table \ref{table3}.  This is $\sim$ 4 times  the M$_{HI}$ $+$ M* with a  ratio of the same order \textcolor{black}{as in  NGC\,4656UV.  We have} no  M$_{dyn}$  for Holmberg IX or Arp\,305.  As for the Arp\,202 TDG candidate, \textcolor{black}{ we  see} no clear indication of the  presence of significant amounts of dark matter. }

\textcolor{black}{Table \ref{table3} shows the estimated extent of each of  the blue candidates, which range from 2.2 kpc to 11 kpc. Because of the wide range in angular scales and distances to the blue candidates it is difficult to properly compare the extent of these  candidates. Moreover there is also a problem of separating the TDG and tail.  We have estimated the extent of the blue candidates from inspection of UV (\textit{GALEX}) images, except for the nearest candidate, Holmberg IX, where we use the value from NED. We concluded that the extent of the blue candidates is typically a few kpc.}



From the FUV {\it(GALEX)}  magnitudes \citep{smith2010} and using the star formation rate (SFR)  vs FUV luminosity relation in \cite{salim}, the estimated SFR of the Arp\,202 TDG is 0.039 M$\odot$yr$^{-1}$. This is lower than the mean value of SFRs found in TDGs \citep{ducmirabel1999}. With its gas reserve, at this SFR, the Arp\,202 TDG can form stars for more than 2 Gyr. Using the H$\alpha$ luminosity quoted in \cite{hancock} and the SFR recipe in \cite{kennicutt}, we also estimated the SFR for Arp\,305 TDG as 0.025 M$_{\odot}$yr$^{-1}$ . The SFR values of  other blue TDG candidates were taken from the literature \citep{ngc4656, sabbi}.  The mean SFR for the blue candidates is $\sim$0.025 M$_{\odot}$yr$^{-1}$, lower than the mean of the \cite{ducmirabel1999} sample. However we note here that the SFR estimates can significantly change depending on the wavebands. This prevents us from suggesting that the SFRs of the blue candidates are actually lower than the TDG average.

\textcolor{black}{The metallicity for the Arp\,202 TDG candidate from \cite{smith2010}  \textcolor{black}{is} log(O/H)+12=8.9, approximately solar metallicity \textcolor{black}{and}  above the log(O/H)+12 mean value of the \cite{ducmirabel1999} sample (8.5). We note here that the  \textcolor{black}{ Arp\,202 TDG candidate's spectrum from which the metallicity is quoted, is} as yet unpublished. Holmberg IX has a similar \textcolor{black}{metalicity} to Arp 202 and the \cite{ducmirabel1999} sample but  NGC\,4656UV is \textcolor{black}{significantly sub--solar}.  Metallicity is one of the strongest \textcolor{black}{tests to distinguish  TDGs  from standard dwarfs. \textcolor{black}{Higher metallicites } are expected if a dwarf formed from metal rich disk material from a parent galaxy as opposed to the poor metalicities for standard dwarfs formed from  pristine gas}. Thus the Arp\,202 \textcolor{black}{candidate's  metallicity provides strong  support for it being a TDG , while} the NGC\, 4656UV metalicity }value raises a question.

\textcolor{black}{ The SDSS \textit{g -- r} colour was calculated using the data from the sources referred to in note j to  Table \ref{table3}. The \textit{g -- r} colour for the Arp\,202 candidate  is 0.21.   As expected from \cite{ngc4656} results, the mean \textit{g--r} colour of the  blue candidates (0.06 $\pm$ 0.13) is bluer than  the  mean  of the 6 TDG candidates in the \cite{smith2010} sample (0.41 $\pm$ 0.20).   Unfortunately this is not directly comparable  to B--V of 0.3 for the 20 TDG candidates in the  \cite{ducmirabel1999}  sample. The  \textit{FUV--g} colour of the Arp\,202 TDG is -0.47,  the bluest  \textit{FUV--g} colour amongst the blue candidates.}

\textcolor{black}{From a study of TDGs, \cite{kaviraj} concluded that 95\% of TDG progenitors are spirals involved in  binary mergers where the parent galaxies' mass ratio  is less than 1:7  (median 1:2.5).  They also concluded TDGs are not produced in interactions where the parent mass ratios exceed 1:11 and  for 95\%  of the  time  the physical separation between the parent galaxies and the TDG is $<$ 20 kpc. These values are  consistent with simulations by \cite{BournaudDuc2006}. For all of the  blue candidates, including the Arp\,202 the progenitor pair  mass ratios are within the  ratio  observed by  \citep{kaviraj}. Table \ref{table3} shows  the protected distance to the nearest gas rich parent galaxy for  three of  the blue candidates is within  20 kpc , but  the Arp\,305 TDG candidate is projected 36 kpc from NGC\,4017.}\\


 \textcolor{black}{Allowing for the small sample size and measurement  uncertainties we see indications that, in addition to the extreme blue colour reported by \cite{ngc4656}, the blue candidates have significant smaller \hi\ masses (2.2$\pm$1.6 $\times$ 10$^{8}$ \msolar)  than the \cite{ducmirabel1999}  sample. It seems likely that the  stellar masses (M*) of the blue candidates are  commensurately smaller  as well. If this is correct, then these candidates would have exceptionally large M$_{HI}$/M* ratios. The lower mean SFR for the blue candidates compared with the  \cite{ducmirabel1999}  sample appears, at least in part, to be  due to the  the smaller \hi\ and stellar masses. }

\begin{table*} 
\centering
\begin{minipage}{190mm}
\caption{Properties of extremely blue TDGs compared to average properties from literature.}
\label{table3}
\begin{tabular}[h]{@{}lllrrrrr@{}}
\hline
Property& \textcolor{black}{Statistic fromTDG } & Arp\,202&NGC\,4656UV&Holmberg IX&Arp\,305  &\\
&\textcolor{black}{candidate samples}&\\
\hline
TDG M*[10$^{8}$ \msolar]\footnote{\textcolor{black}{ M* \textcolor{black}{ for Arp\,202 and NGC\,4656UV  are estimated from SDSS \textit{g} and\textit{ r} band colours \textcolor{black}{ following log(M*/L$_{g}$)=a$_{g}$+b$_{g}$ $\times$ (g-r) } \citep{Bell2003} and g-- band solar luminosity from parameter from \cite{Blanton}, using colours from  \cite{smith2010} and \cite{ngc4656} respectively. For Holmberg IX M* is from \cite{sabbi} and M* for Arp\,305  is} from \cite{hancock} }}&&\textcolor{black}{0.2}&\textcolor{black}{0.4} &\textcolor{black}{0.02}&\textcolor{black}{0.04}\\
TDG M$_{HI}$ [10$^{8}$ \msolar]\footnote{\textcolor{black}{M$_{HI}$ for: \textcolor{black}{Arp\,202 fromTable \ref{table2},}  NGC\,4656UV from \cite{ngc4656}, Holmberg IX from Swaters \& Balcells (2002) and Arp\,305 van Moorsel 1983.}}& 16.0\footnote{Duc \& Mirabel, 1999} &1.0 &3.8&3.3&0.6\\
TDG Mdyn [10$^{8}$ \msolar]\footnote{\textcolor{black}{M$_dyn$ for Arp\,202 = 3.39 $\times$ 10$^4$ $\times$ a$_{HI}$ $\times$ d $\times$ (W$_{20}$/2), where a$_{HI}$ = diameter in arcmin, d = the distance to the object and W$_{20}$ = the line width at 20\% of the peak \citep{haynes}. The H{\sc i} beam major axis of the high-resolution map was used as a$_{HI}$ and since the signal is present in only two channels in our high-resolution cube, the net width of the two channels was used as W$_{20}$.  NGC\,4656UV from \cite{ngc4656} }} &&3.9  &\textcolor{black}{19}& --& --\\
TDG extent [kpc]\footnote{\textcolor{black}{Extents were  measured from UV  \textit{(GALEX)}  images  except for and Holmberg IX which is from NED. }}&&\textcolor{black}{4.0} &\textcolor{black}{6.3}&2.25&\textcolor{black}{11.0}\\
TDG SFR [ 10$^{-3}$ \msolar/yr ]\footnote{\textcolor{black}{SFR for the Arp\,202 TDG candidate is estimated using the FUV luminosity quoted in \cite{smith2010} and the SFR recipe in \cite{salim},  NGC\,4656UV is from \cite{ngc4656},  Holmberg IX from \cite{sabbi} and  SFR for Arp\,305 is estimated using the H$\alpha$ luminosity quoted in \cite{hancock} and the SFR recipe in \cite{kennicutt} }}&79\footnote{Duc \& Mirabel, 1999}&39&27&\textcolor{black}{8}&25\\
TDG Metallicity \footnote{\textcolor{black}{Metallicity for : Arp\,202 is from Smith et al. (2010a),  NGC\,4656UV is from \cite{ngc4656}, Holmberg IX is from \cite{Makarova}}}  & 8.5\footnote{Duc \& Mirabel, 1999} &8.9 &\textcolor{black}{-1.7}& 8.5& --\\
&(12+log(O/H)) & (12+log(O/H)) & [Fe/H]& (12+log(O/H)) \\
TDG colour\footnote{\textcolor{black}{g--r color from \cite{smith2010}, except NGC\,4656UV  which is converted from Table 2 of \cite{ngc4656}}}&0.30\footnote{Duc \& Mirabel, 1999}&0.21&\textcolor{black}{-- 0.02}&\textcolor{black}{-- 0.07}&0.13\\
   \,\,\,\, \textcolor{black}{-- colour system}&(B-V)& (g--r) & (g--r) & (\textcolor{black}{g--r}) & (g--r)  \\
TDG FUV--g\footnote{\textcolor{black}{FUV--g colour from \cite{smith2010}, except NGC\,4657UV  which is converted from Table 2 of \cite{ngc4656}}} &--&-- 0.47& \textcolor{black}{0.64}&\textcolor{black}{-- 0.3} &0.24\\
Parent pair mass ratio&$>1:7$, median = 2.5\footnote{\textcolor{black}{ \cite{kaviraj}}}&1:1.4&\textcolor{black}{1:2.6}&\textcolor{black}{1:1.5}&1:2.9\\
TDG \textcolor{black}{projected} distance from parent (kpc)\footnote{\textcolor{black}{The \textcolor{black}{projected} distances are: for Arp\,202 projected distance from \textcolor{black}{NGC\,2719A}, for NGC\,4656UV  \cite{ngc4656}, \textcolor{black}{for Holmberg IX, projected distance from M81 from NED}  and Arp\,305 projected distance from the position of NGC\, 4017 per Hancock et al.2009 }}&$<$20\footnote{\textcolor{black}{ \cite{kaviraj}}}&\textcolor{black}{18} &11&\textcolor{black}{13}&36\\
\textcolor{black}{TDG estimated age [Myr]\footnote{\textcolor{black}{The age estimates are: NGC\,4656UV from \cite{ngc4656} and  Holmberg XI from \cite{sabbi} }}}&--&--&\textcolor{black}{292}&\textcolor{black}{200}& --\\

\hline
\end{tabular}
\end{minipage}
\end{table*}

\section{Summary and concluding remarks}
\label{summary}
\textcolor{black}{Our  GMRT \hi\  morphology and kinematic results clearly link the \hi\ tidal tail and the \hi\ TDG counterpart to the interaction between  NGC\,2719 and NGC\,2719A. This and the similarity of the Arp\,202 candidate's properties to other TDG candidates with extreme blue colours (Table \ref{table3}), and in particular Holmberg IX, strengthens the case for the Arp\,202 candidate being a TDG formed under  the standard TDG formation scenario.  Observations to estimate  molecular gas content and kinematics as well as detailed stellar population analysis could further improve the understanding of the origin of this  TDG candidate. }

\textcolor{black}{We compared properties of the Arp\,202 candidate and three \textcolor{black}{other} extremely blue TDG candidates identified by \cite{ngc4656}  to  larger samples of TDG candidates. \textcolor{black}{All four of these} extremely blue candidates have \hi\ (and probably M*) masses \textcolor{black}{comparable} to the smallest TDGs found in the literature. These lower masses are probably at least partially responsible for the low SFR  compared to the   \cite{ducmirabel1999} TDG sample. All of the extremely blue candidates are H{\sc i} rich, with the M$_{HI}$/M* ratios ranging approximately between 5 and 150.  \textcolor{black}{The number of  extremely blue TDG candidates examined  is too small to conclude whether or not they are a distinct subgroup with TDGs. }  }

\section{Acknowledgments}

We thank the staff of the {\it GMRT} who have made these observations possible. The {\it GMRT} is operated by 
the National Centre for Radio Astrophysics of the Tata Institute of Fundamental Research. This research 
has made use of the NASA/IPAC Extragalactic Database (NED) which is operated by the Jet Propulsion Laboratory, 
California Institute of Technology, under contract with the National Aeronautics and Space Administration. CS thanks Dr. Yongbeom Kang from KASI for useful discussions.



\onecolumn

\end{document}